\documentclass[a4paper]{jpconf}

\usepackage{graphicx}

\newcommand{\tsol}{\ensuremath{\theta_{12}}}
\newcommand{\tatm}{\ensuremath{\theta_{23}}}
\newcommand{\trea}{\ensuremath{\theta_{13}}}
\newcommand{\sst}{\ensuremath{\sin^2 2 \theta_{13}}}
\newcommand{\dcp}{\ensuremath{\delta_{\mathrm{{CP}}}}}

\newcommand{\dmsol}{\ensuremath{\Delta m^2_{21}}}
\newcommand{\dmatm}{\ensuremath{\Delta m^2_{31}}}

\newcommand{\gev}{\ensuremath{\mathrm{G}e\mathrm{V}}}
\newcommand{\mev}{\ensuremath{\mathrm{M}e\mathrm{V}}}
\newcommand{\ev}{\ensuremath{e\mathrm{V}}}
\newcommand{\gwt}{\ensuremath{\mathrm{GW_\mathrm{th}}}}

\newcommand{\uttf}{\ensuremath{{}^{235}\mathrm{U}}}
\newcommand{\utte}{\ensuremath{{}^{238}\mathrm{U}}}
\newcommand{\pttn}{\ensuremath{{}^{239}\mathrm{Pu}}}
\newcommand{\ptfo}{\ensuremath{{}^{241}\mathrm{Pu}}}

\newcommand{\minib}{MiniBooNE}

\begin{document}

\title{Status and perspectives of short baseline studies}

\author{Mark Dierckxsens}

\address{Department of Physics, University of Chicago, Chicago, IL 60637, U.S.A.\\
Now at: Universit\'e Libre de Bruxelles, Science Faculty CP230, B-1050 Brussels, Belgium}

\ead{mdier@hep.uchicago.edu}

\begin{abstract}

The study of flavor changing neutrinos is a very 
active field of research. 
I will discuss the status of ongoing and near term 
experiments investigating neutrino properties at short distances 
from the source.
In the next few years, the Double Chooz, RENO and Daya Bay 
reactor neutrino experiments will start looking for signatures 
of a non-zero value of the mixing angle $\theta_{13}$ with much 
improved sensitivities.
The \minib\ experiment is investigating the LSND anomaly by looking
at both the $\nu_{\mu} \rightarrow \nu_{e}$ and 
$\bar{\nu}_{\mu} \rightarrow \bar{\nu}_{e}$ appearance channels.
Recent results on cross section measurements will be 
discussed briefly.

\end{abstract}

\section{Introduction}

The relation between the flavor and mass eigenstates of the neutrinos 
can be described by the PMNS matrix \cite{Maki:1962mu,Pontecorvo:1967fh}.
In case of 3-neutrino mixing, the parameters relevant to oscillations 
describing this matrix are
three mixing angles, \tsol, \tatm\ and \trea,  
and a CP-violating phase \dcp.
The frequencies of oscillations between the different states are governed
by the energy, distance and the two possible differences between
the masses squared, \dmsol\ and \dmatm.

Since the confirmation of neutrino oscillations a decade ago 
\cite{Fukuda:1998mi}, great progress has been made in the measurements
of these parameters.
Two of the mixing angles, \tsol\ and \tatm, and the size of the two mass
splittings, \dmsol\ and \dmatm, are well measured
(for a recent combination of global data, see e.g.~\cite{Fogli:2009zz}).
The most stringent limit on the third angle comes from the 
CHOOZ reactor neutrino experiment \cite{Apollonio:2002gd}: 
$\sin^2 2 \theta_{13} < 0.16~(90\%\ \mathrm{C.L.})$ at 
$\dmatm = 2.5\ 10^{-3} \ev$.
However, small contributions from various experiments add up
to prefer a non-zero value of $\theta_{13} \simeq 8^{o}$ at 
the $2 \sigma$ level \cite{Fogli:2009zz,palazzo}.
Several experiments currently under construction will probe 
well into this region.
Those observing neutrinos at nuclear reactors will be covered 
in the next section.
%
%

However, not all measurements fit clearly into the above picture. 
The LSND experiment has observed an excess 
of $87.9 \pm 23.2\ (3.8 \sigma)$ events above background
in the $\bar{\nu}_{\mu} \rightarrow \bar{\nu}_{e}$ appearance search 
\cite{Aguilar:2001ty}.
The L/E value requires 
a mass splitting different than the solar and atmospheric 
$\Delta m^2$ to explain this effect with neutrino oscillations.
Three mass differences need mixing with a unobserved fourth type 
of neutrino with different mass, suggesting the existence of
sterile neutrinos.
The \minib\ experiment was designed to investigate this
excess in detail, and will be covered in Section~\ref{sec:miniboone}.

\section{Reactor neutrino experiments}

\subsection{Neutrino oscillations at nuclear reactors}

Nuclear reactors are a pure and isotropic source of anti-electron neutrinos
produced at a rate of about $2 \cdot 10^{20} / \gwt /s$. 
The particles are produced through the decay of the fission products in the
decay chain of predominantly \uttf, \pttn, \ptfo\ and \utte.
The $\beta$-spectra from the first three isotopes are measured to a precision 
of $1.8 \%$ \cite{Schreckenbach:1985ep,Hahn:1989zr}. 
These spectra are converted into a predicted neutrino spectrum 
resulting in a final error of $2.5-4\%$ depending on the energy.
Plans are underway to decrease this error significantly by measuring 
the \utte\ $\beta$-spectrum and performing improved conversion calculations
including more branches in the decay chain.

The survival probability for anti-electron neutrinos is given by
\begin{equation}
  \label{eq:panue}
  P(\bar{\nu}_e \rightarrow \bar{\nu}_e) \simeq
  1 - \sin^2 2 \theta_{13} \sin^2(\frac{1.27 L \dmatm}{E})
  - \cos^4 \theta_{13} \sin^2 2 \theta_{12} \sin^2 (\frac{1.27 L \dmsol}{E}),
\end{equation}
with $L$ measured in m, $E$ in \mev\ and $\Delta m^2$ in \ev.
In case of a non-zero $\theta_{13}$, the first oscillation node is 
around 2~km for typical energies in nuclear reactor experiments.
An observation of neutrino disappearance at this distance will
constitute a clean measurement of $\sin^2 2 \theta_{13}$:
the last term containing the solar parameters can be neglected and the 
probability only depends on \dmatm, a well measured quantity.
The probability is not affected by matter effects due to the 
short distances and is insensitive to the CP-violating phase.
Several experiments have searched unsuccessfully for the disappearance
of rector neutrinos at various distances due to a non-zero \trea.
The best limit was set by the Chooz experiment to be 
$\sin^2 2 \theta_{13} < 0.16 \ (90 \%\ \mathrm{C.L.})$ 
\cite{Apollonio:2002gd}.

Three experiments are currently under construction with a much 
improved sensitivity: Double Chooz, RENO and 
Daya Bay.
The statistical uncertainty will be lowered by building larger
detectors, running for a longer time and by being located near
more powerful reactors.
The reduction in systematics uncertainty is mainly obtained by placing one 
or more identical detectors close to the reactor cores.
A ratio measurement will eliminate the cross section,
the neutrino flux and some detector uncertainties.
There is also a gain from an improved detector design 
with lower thresholds,
increased efficiencies and implementation of a very detailed calibration
program.
Further enhancements are obtained by lowering the background
contamination with better veto systems, shielding, higher 
overburden and the use of materials with improved radio-purity.

\subsection{Detection technique} \label{ssec:dettech}

The anti-electron neutrinos are measured through the inverse $\beta$-decay
process $\bar{\nu}_e + p \rightarrow e^{+} + n$,
with a threshold energy of 1.8 \mev. 
Folding the neutrino flux from the reactor with the cross section results
in an observable spectrum peaking just below 4~\mev.

The three experiments  employ similar detector designs
with three regions containing different types of liquids, 
separated by optically transparent acrylic vessels.
The central region is filled
with liquid scintillator doped with a small amount of Gadolinium, 
typical $0.1\%$ in weight.
The produced positron annihilates instantly
and deposits a total amount of visible energy 
closely related to the neutrino energy: 
$E_{\mathrm{vis}} \simeq E_{\nu} - 0.8~\mev$.
The neutron slows down through thermalization before being
captured by a nucleus. 
This will most often happen on Gadolimium 
due to its high cross section for neutron capture.
The typical capture time is $30 \mu s$ and releases a total amount 
of energy around 8~\mev\ in the form of several gammas. 
The target region is surrounded by the gamma catcher: a volume 
with undoped scintillator to capture any gamma escaping the target. 
The outside region contains the photo-multiplier tubes (PMT) and 
is filled with non-scintillating oil.
This design suppresses a large fraction of the radioactivity coming
from the PMTs and surrounding material.
The detectors are also surrounded by active veto detectors and
passive shielding to further suppress natural radioactivity and 
backgrounds related to cosmic muon interactions.  

The typical signature of a prompt signal plus a delayed energy deposit 
consistent with neutron capture on Gadolinium is used 
to select neutrino candidates.
The background events can be categorized into two classes. 
The accidental backgrounds consist of a positron-like signal from 
natural radioactivity present in the PMTs, the detector materials and
the surrounding rock, followed close in time by an independent 
neutron-like signal. 
These can be actual neutrons generated by cosmic muons
interacting in the material surrounding the detector, or high energy
gammas mimicking a neutron capture on Gadolinium.
The correlated backgrounds are 
produced by one single event with multiple energy deposits.
A fast neutron produced by cosmic muons
slows down by recoiling on the protons in the 
scintillator producing electron-like signals until it 
gets captured by the Gadolinium.
Other sources are long lived isotopes produced in high energetic
showering cosmic muon interactions undergoing a $\beta$-n decay.
The isotopes have typical livetimes of $\mathcal{O}(100 \mathrm{ms})$ 
and are impossible to eliminate by vetoing every cosmic muon.
The largest contribution comes from ${}^9$Li, but other isotopes like
${}^8$He and ${}^{11}$Li also contribute.

\subsection{Future experiments}

\begin{table}[tb]
  \centering
  \begin{tabular}{|c|cccc|ccc|}
    \cline{2-8}
    \multicolumn{1}{c|}{} & Power   & $\mathrm{L_{Near}}$ & $\mathrm{L_{Far}}$ 
    & $\mathrm{M_{target}}$
    & $\sigma_{\mathrm{stat}}$      & $\sigma_{\mathrm{syst}}$ 
    & $\sin^2 2 \theta_{13} > $  \\
    \multicolumn{1}{c|}{} &  [\gwt] & [m] & [m] & [ton] & [$\%$] & [$\%$] & ($90 \%$ C.L.)\\
    \hline
    Double Chooz &  8.6 & 400  & 1050 & 8.3 & 0.5 & 0.6 & 0.03 \\
    RENO         & 17.3 & 290  & 1380 & 16  & 0.3 & 0.5 & 0.02 \\
    Daya Bay     & 17.4 & 360 (500)  & 1990 (1620) & 80  & 0.2 & 0.4 & 0.01 \\
    \hline
  \end{tabular}
  \caption{Total reactor power, distance $L$ to near and far detectors 
    and far detector target mass $\mathrm{M_{target}}$ of the
    Double Chooz, RENO and Daya Bay experiments. 
    The distance for Daya Bay is to the DB plant, 
    the ones between brackets to the LA and LA II plants.
    The expected statistical and systemic uncertainties and sensitivity to 
    $\sin^2 2 \theta_{13}$ at 90$\%$ confidence level are given on the right.}
  \label{tab:reacexps}
\end{table}

The three experiments currently under construction are 
described below and summarized in Table~\ref{tab:reacexps}. 
The expected sensitivities mentioned are always upper limits on  
$\sin^2 2 \theta_{13}$ at $90\%$ confidence level in case no 
signal is observed.

\paragraph{Double Chooz.}

The experiment \cite{Ardellier:2006mn} is located near 
the Chooz-B Power Plant in Northern France, which consists of two 
reactor cores operating at a total power of $8.6~\gwt$.
The near detector will be located at a distance of about 400~m from 
the cores.
The far detector is being constructed in the same hall as
the CHOOZ experiment at a distance of 1050~m.
The target regions of both detectors will be 8.3~ton.

\begin{figure} 
\begin{minipage}{0.35\textwidth}
\includegraphics[width=\textwidth]{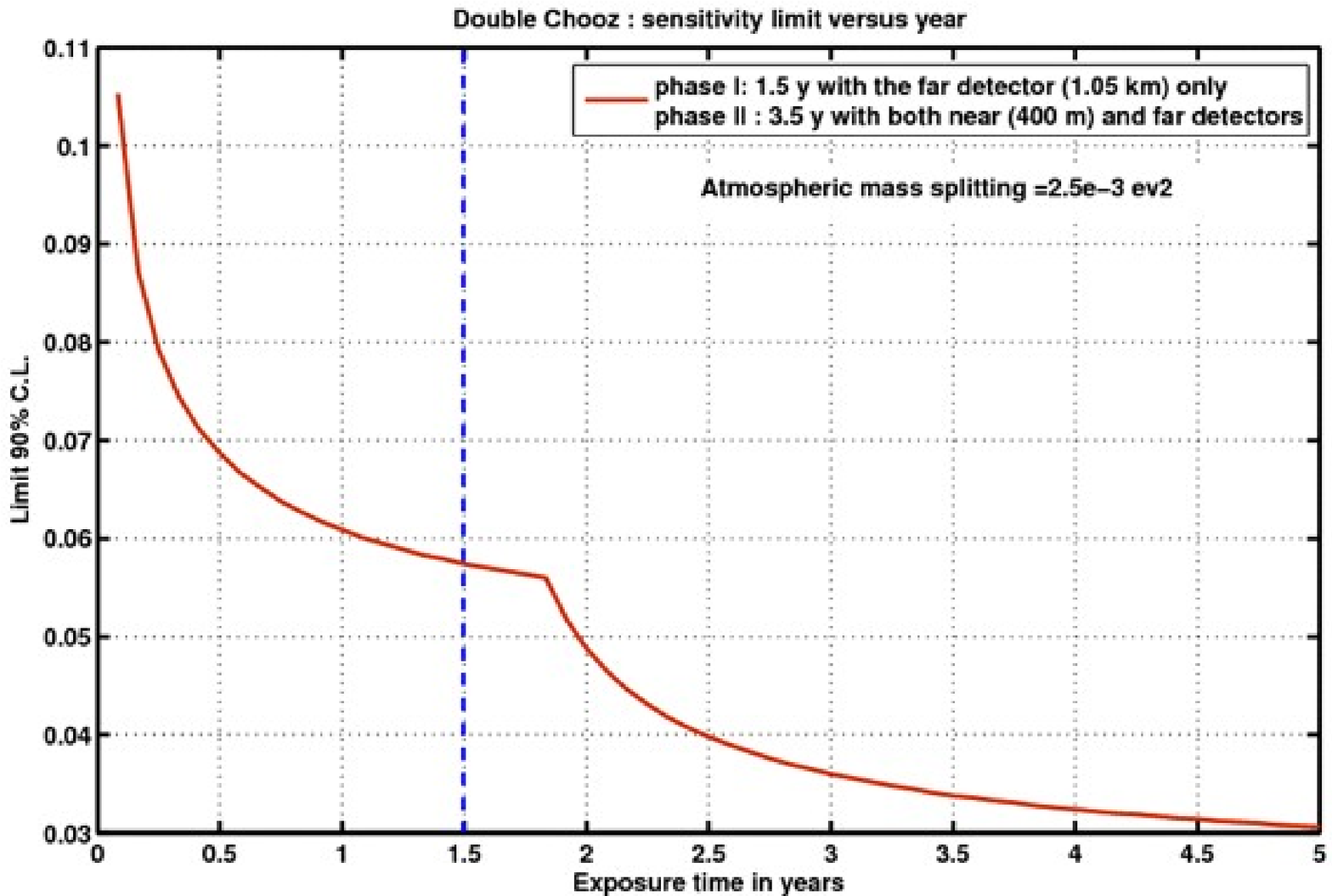}
\vspace*{0.1cm}
\end{minipage}
\hspace{0.01\textwidth}
\begin{minipage}{0.32\textwidth}
\includegraphics[width=\textwidth]{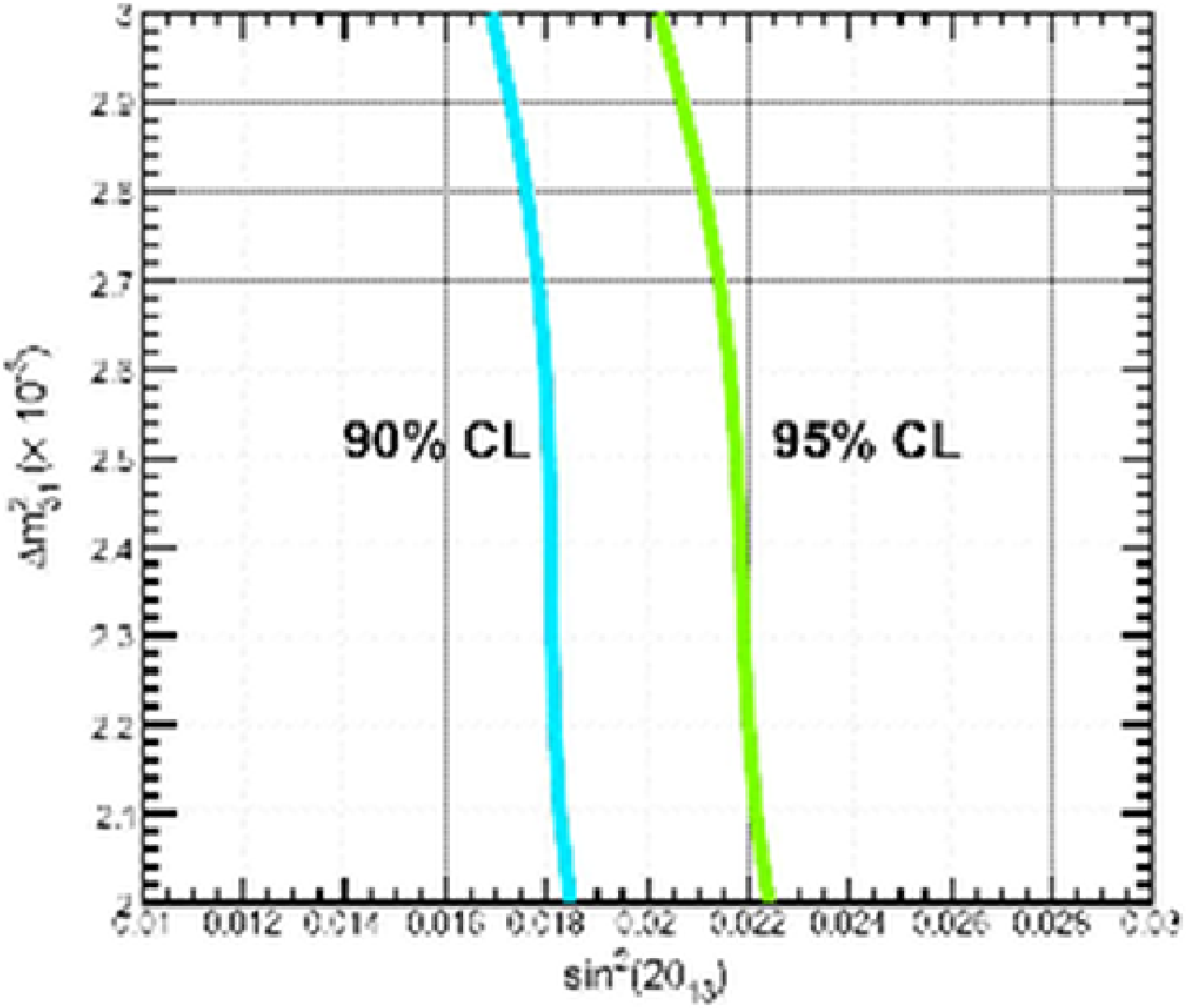}
\end{minipage} 
\begin{minipage}{0.30\textwidth}
\vspace{0.1cm}
\includegraphics[width=\textwidth]{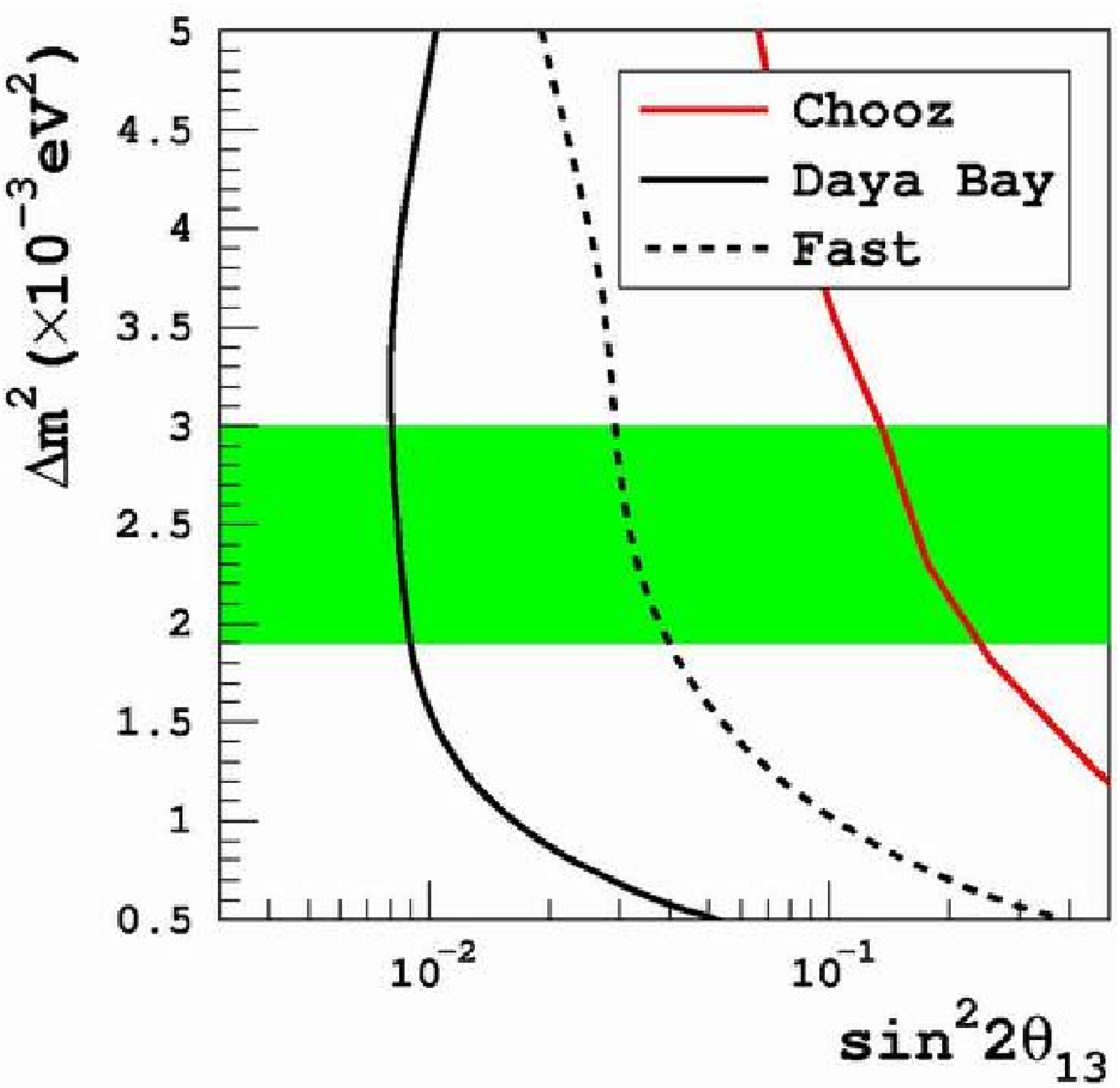}
\end{minipage}

\caption{The expected sensitivity to \sst\ at $90\%$ C.L.: 
as a function of running time for Double Chooz (left) where the 
near detector becomes operational after 1.5 years of far detector
only running; 
as function of \dmatm\ for RENO (middle) and Daya Baye (right) after 
three years of data taking with all detectors and reactors fully operational.}
\label{fig:reac}
\end{figure}

The far detector is expected to be operational by spring 2010 for
the first phase with far detector only running.
After one year, the expected sensitivity to \sst\ will be 0.06.
The second phase of the experiment with both detectors operational
will start in 2011, with a sensitivity to \sst\ of 0.03 after three
years of data taking.
The expected sensitivity over time can be seen in the left plot of 
Figure~\ref{fig:reac}.
More details about and the status of the experiment are covered 
in \cite{maricic}.

\paragraph{RENO.} 

The experiment  \cite{Kim:2008zzb} is being built near the YeongGwang 
power plant in the south-west of South Korea.
The 6 operational cores lie on one axis and produce a total power of
$17.3~\gwt$. 
The target regions of both detectors will be 20 ton.
The near detector hall is located at 290~m from the closest point
to the reactor cores axis and
the far detector hall is at a distance of 1380~m.

Both detectors will be ready for data taking during the early
months of 2010.
%
The expected sensitivity to \sst\ after 3 years will be 0.02, and 
can be seen as function of \dmatm\ in the middle plot of 
Figure~\ref{fig:reac}.

\paragraph{Daya Bay.}
The experiment \cite{Guo:2007ug} will use neutrinos produced by the 
Daya Bay (DB) and Ling Ao (LA) Power plants located in just north-east 
of Hong Kong.
In 2011, the LA II power plant will be operational
resulting in a total power of 17.4~\gwt\ produced by six
cores.
A total of 4 identical modules measuring 20 ton each will be placed at the far
hall, 1990~m and 1620~m away from the DB and LA reactors, respectively.
Two modules will be placed at a distance of about 360m from the DB plant, 
and another two at about 500m from the LA plants. 

The far hall will be ready for data taking in Summer 2011. 
After 3 years with all detector modules operational and 
6 reactor cores at full power, 
the sensitivity to \sst\ is expected to reach 0.01, which can be seen from 
the right plot in Figure~\ref{fig:reac}.
More details about and the status of the experiment are covered in
\cite{Heeger}.


\section{The \minib\ experiment} \label{sec:miniboone}

\subsection{Experimental setup}

The \minib\ experiment \cite{AguilarArevalo:2008qa} is designed to 
address $\nu_e$ appearance in a $\nu_{\mu}$ beam with the same 
L/E value as LSND but 
under different experimental conditions. 
A 8.9~\gev\ proton beam from the Booster at FNAL hits a beryllium 
target.
The produced hadrons are focused in the forward direction
by a magnetic horn. 
They are allowed to decay in a tunnel followed by dirt
to absorb all particles but neutrinos. 
This results in a $\nu_{\mu}$ beam peaked around 700~\mev.
The polarity of the horn can be reversed to select a $\bar{\nu}_{\mu}$
beam.
An 800 ton mineral oil \v{C}erenkov detector is placed at about 
500~m downstream the neutrino beam. 
A short overview of the results are given in the next section. 
More details are covered in \cite{Shaevitz}.

\subsection{Results}

\begin{figure}
\begin{minipage}{0.31\textwidth}
\includegraphics[width=\textwidth]{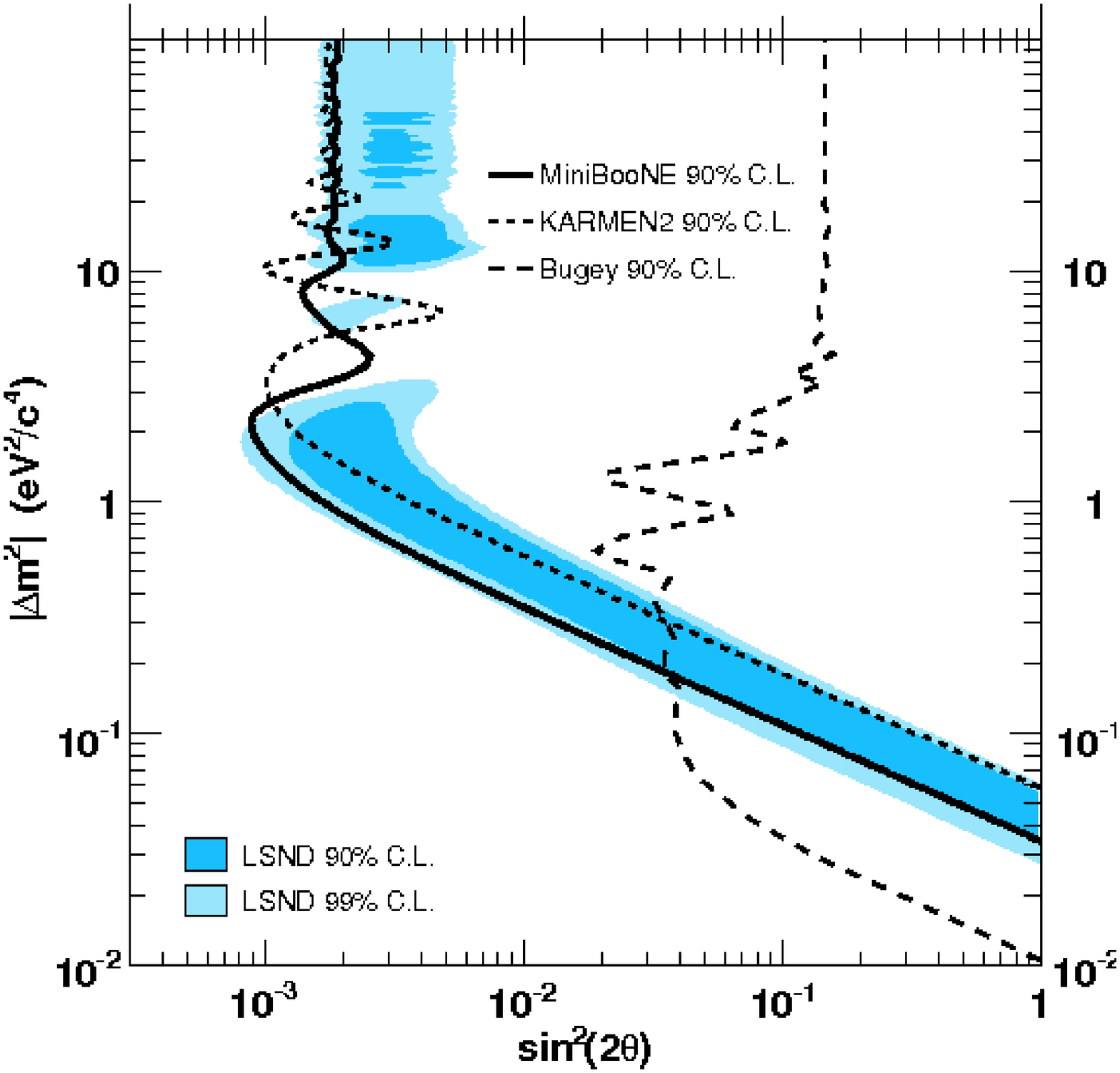}
\end{minipage}
\hspace{0.01\textwidth}
\begin{minipage}{0.31\textwidth}
\includegraphics[width=\textwidth]{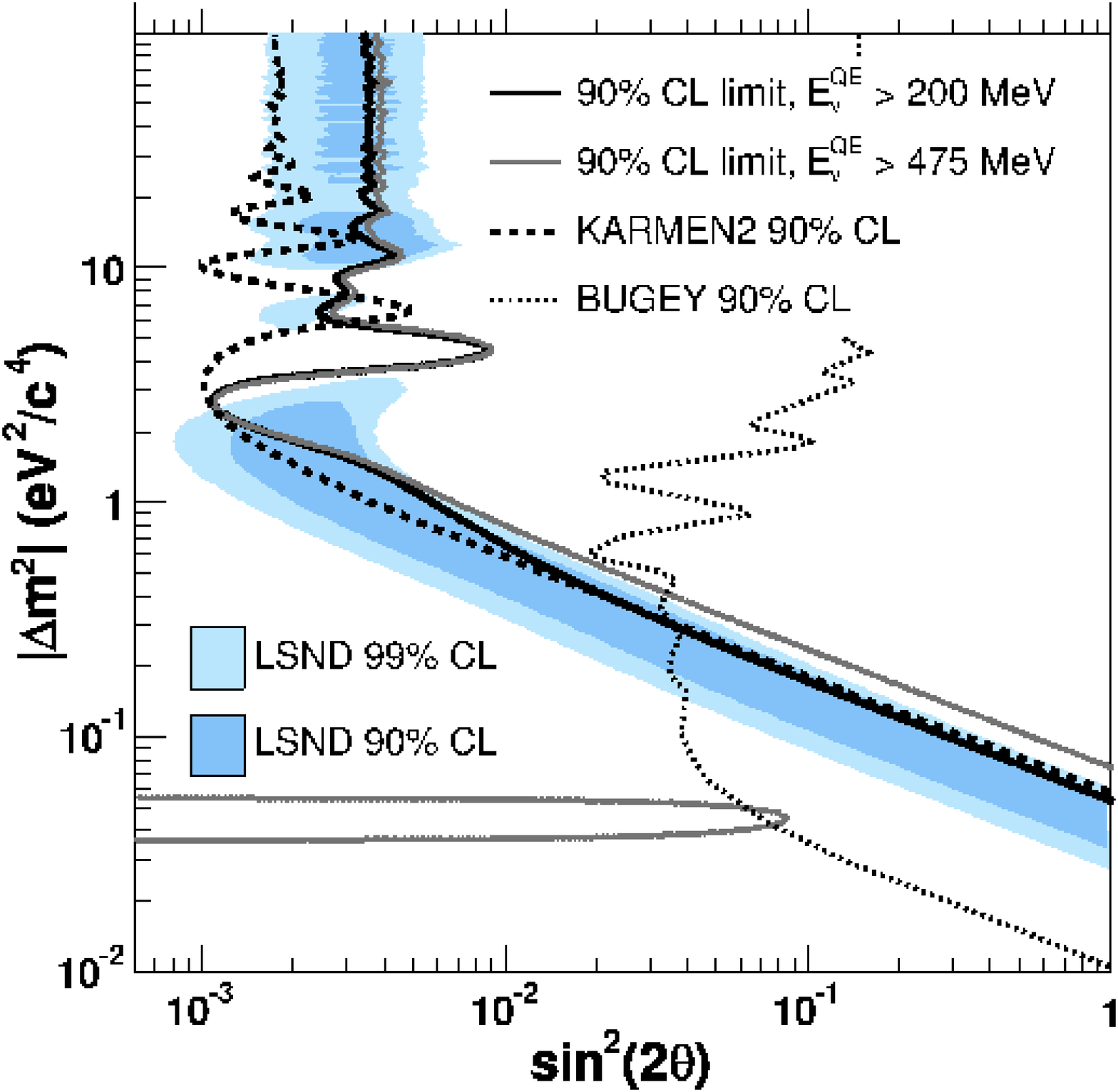}
\end{minipage} 
\begin{minipage}{0.35\textwidth}
\includegraphics[width=\textwidth]{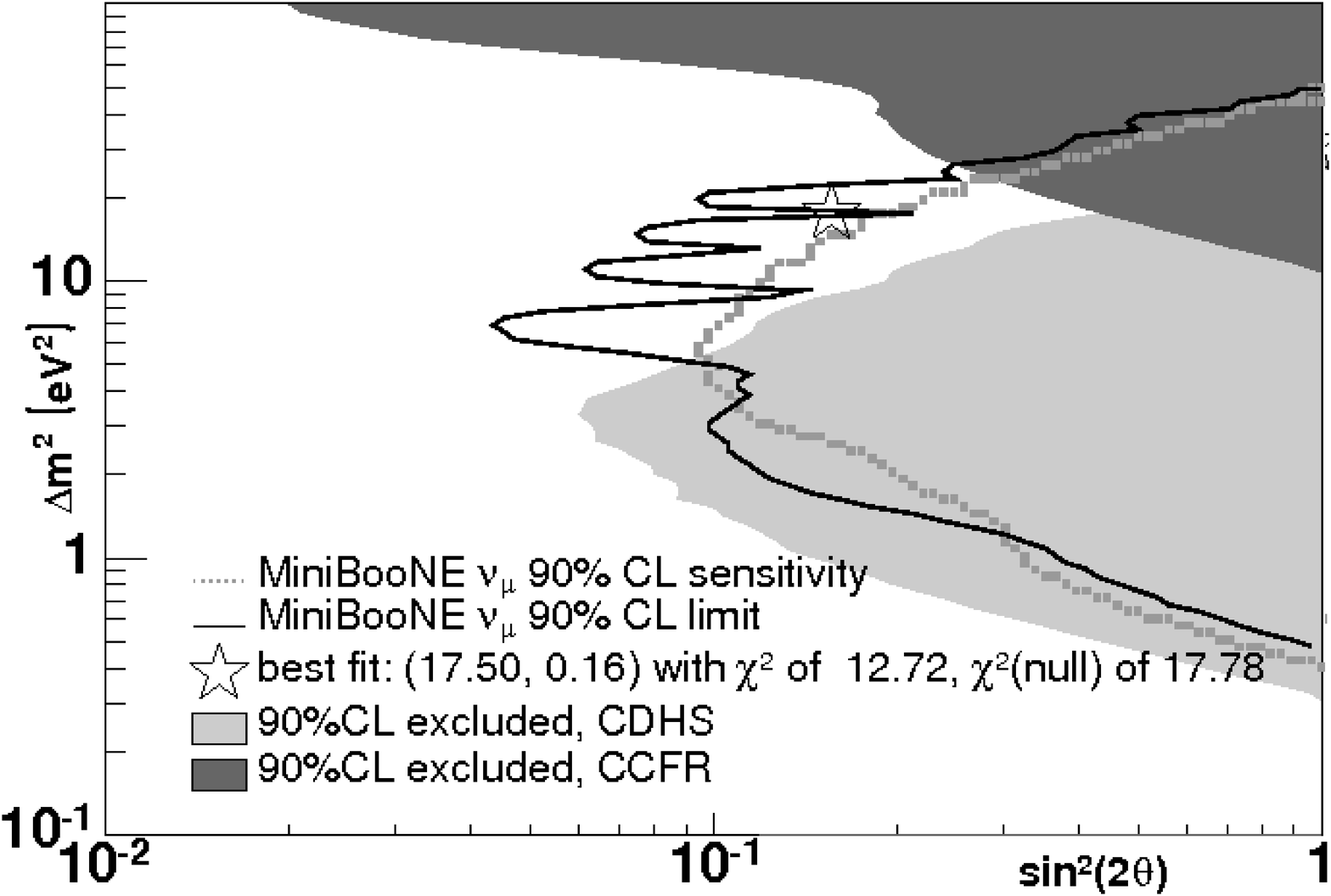}
\end{minipage}
\caption{The observed \minib\ exclusion limits on
$\nu_{\mu} \rightarrow \nu_{e}$ (left) and 
$\bar{\nu}_{\mu} \rightarrow \bar{\nu}_{e}$ (middle) appearance
and on $\nu_{\mu}$ disappearance (right).}
\label{fig:minib}
\end{figure}

\paragraph{$\nu_{\mu} \rightarrow \nu_{e}$ appearance.}
The full data sample of $6.5 \cdot 10^{20}$ protons-on-target (PoT) taken
in neutrino mode is analyzed, selecting 
charged current quasi-elastic $\nu_{e}$ interactions.
No excess of events was observed in the signal region above 475~\mev\
corresponding to the LSND results \cite{AguilarArevalo:2007it}. 
This rules out the 2-neutrino oscillation hypothesis as source of 
the LSND excess (assuming no CP \& CPT violation). 
The resulting exclusion limit can be seen in 
the left plot of Figure~\ref{fig:minib}.

However, an excess of $129\pm43$ events was observed above
background at energies between 200 and 475~\mev\
\cite{AguilarArevalo:2008rc}.
The size of the signal is  similar to the LSND result
but the shape of the energy spectrum is not consistent with
2 neutrino oscillations.
The effect is also observed looking at far off-axis neutrinos 
coming from the NuMI beam \cite{Adamson:2008qj}. An improvement 
in the latter analysis is expected, including a larger data sample 
and reduced systematic uncertainties.

\paragraph{$\bar{\nu}_{\mu} \rightarrow \bar{\nu}_{e}$ appearance.}

The anti-neutrino mode is a direct test of the LSND result. 
Results from a limited data sample of $3.4 \cdot 10^{20}$ PoT show no 
significant excess over background, 
neither in the LSND region, nor at lower energies 
\cite{AguilarArevalo:2009xn}.
The obtained exclusion limits are show in the middle plot
of Figure~\ref{fig:minib}. 
In total $5.1 \cdot 10^{20}$ PoT are recorded in anti-neutrino mode
and $5 \cdot 10^{20}$ PoT more are expected.
Such a threefold increase in the data sample should shine more light on
these inconclusive results.

\paragraph{$\nu_{\mu}$ and $\bar{\nu}_{\mu}$ dispearance.}

Looking at this channel set limits in previously unexplored parameter
space \cite{AguilarArevalo:2009yj}, which can be seen in the right plot
of Figure~\ref{fig:minib} for the $\nu_{\mu}$ channel.
The results will be updated including the data from the SciBooNE experiment,
a fine-grained tracking detector located about 100m from the target
and which took data from June 2007 until August 2008.

\subsection{Beyond \minib}

The MicroBooNE collaboration \cite{microboone} proposes to build a 70 ton 
Liquid Argon TPC near \minib\ as an advanced R\&D project
and to investigate the observed low energy excess.
If all funding is assured, the project can start to take data
as early as 2011. 
Currently, a 170L LAr TPC, called ArgoNeut, is taking data in the 
NuMI beam in front of the MINOS detector \cite{Spitz}.

OscSNS \cite{oscsns} is a proposal
to place a \minib-like detector at 60m from the target
of the Spalation Neutron Source at ORNL,
a 1~\gev, pulsed (60~Hz), 1.4~MW proton beam produces $\pi^+$s which
decay at rest. 
The experiment is expected to have a 15 times better sensitivity than
the LSND experiment.

\section{Cross section measurements}

The intense neutrino beams also allow for much improved cross 
section measurements.
These are in particular important for future neutrino oscillation
experiments operating in the few \gev\ range.
The currently ongoing experiments are \minib\ and SciBooNE 
\cite{AguilarArevalo:2006se} in the
Booster neutrino beam (0.4~-~2~\gev) and MINOS 
\cite{Michael:2008bc} in the NuMI beam (1~-~20~\gev).
One of the many updated and new measurements 
reported on at the recent NuInt09 workshop \cite{nuint}
is the quasi-elastic (CCQE) cross section.
All ongoing experiments prefer a value of $M_{A}$ around 1.35~\gev 
\cite{AlcarazAunion:2009ku,Katori:2009du,Dorman},
where $M_A$ is the axial form factor in the relativistic Fermi Gas Model.
These measurements on carbon and iron are higher than previous results
on $\mathrm{D}_2$, and on carbon at higher energies by NOMAD 
\cite{Lyubushkin:2008pe}.
The MINERvA experiment \cite{Drakoulakos:2004gn} will be able to 
study these discrepancies in much more detail when it starts
operating early 2010.
This dedicated neutrino scattering
experiment, located in front of the MINOS near detector,
has the ability to measure cross sections on different target materials 
in the energy range of 1~-~20~\gev.
Also the T2K \cite{Itow:2001ee} and NOvA \cite{Ayres:2004js} near 
detectors will provide valuable cross section measurements around 
0.8~\gev\ and 2~\gev, respectively.


\section*{References}

\end{document}